\documentclass[12pt]{iopart}
\usepackage[dvips]{graphics}
\usepackage{graphicx}
\usepackage{amsfonts}
\usepackage{amssymb}
\usepackage{subfigure}
\usepackage{cite}

\begin{document}

\title[Dark-bright solitons in Bose-Einstein condensates at finite temperatures]
{Dark-bright solitons in Bose-Einstein condensates at finite temperatures}

\author{V.~Achilleos$^1$, D.~Yan$^2$, P.G. Kevrekidis$^2$, and D.J. Frantzeskakis$^1$}

\address{$^1$
Department of Physics, University of Athens, Panepistimiopolis, Zografos, Athens 15784, Greece}

\address{$^2$
Department of Mathematics and Statistics, University of Massachusetts, Amherst, MA 01003-4515 USA}

\begin{abstract}

We study the dynamics of dark-bright solitons in binary mixtures of Bose gases at finite temperature using a system of two coupled dissipative Gross-Pitaevskii equations. We develop a perturbation theory for the two-component system to derive an equation of motion for the soliton centers and identify different temperature-dependent damping regimes. We show that the effect of the bright (``filling'') soliton component is to partially stabilize ``bare'' dark solitons against temperature-induced dissipation, thus providing longer lifetimes. We also study analytically thermal effects on dark-bright soliton ``molecules'' (i.e., two in- and out-of-phase dark-bright solitons), showing that they undergo expanding oscillations while interacting. Our analytical findings are in good agreement with results obtained via a Bogoliubov-de Gennes analysis and direct numerical simulations.

\end{abstract}

{\tableofcontents}

\maketitle

\pagestyle{plain}

\section{Introduction}

Macroscopic nonlinear excitations of atomic Bose-Einstein condensates (BECs) \cite{book2a,book2} have been a
subject of intense theoretical and experimental research over the last
few years \cite{BECBOOK}. More specifically,
matter-wave {\it dark} and {\it bright} solitons, that can be formed in single-component BECs with repulsive
or attractive interatomic interactions respectively, have been observed in a series of experiments while
their statics and dynamics have been extensively studied theoretically in various settings (see, e.g.,
\cite{rab,revnonlin,djf} for recent reviews). Of particular interest are {\it coupled} dark-bright (DB) solitons
that may exist in binary mixtures of BECs with repulsive interatomic interactions (such as ones composed
by different hyperfine states of $^{87}$Rb atoms \cite{binary,mertes}): these solitons are frequently called
{\it symbiotic} ones, as the bright soliton component (which does not exist in the system with repulsive
interactions \cite{rab}) can be supported due to the nonlinear coupling with the dark soliton component. Such structures
have recently been observed experimentally in a $^{87}$Rb BEC mixture using a phase-imprinting method
\cite{hamburg} or in two counter-flowing $^{87}$Rb BECs \cite{pea,peb}, while they have also been studied in
various theoretical works in continuum \cite{pea,peb,BA01,the} and
discrete~\cite{azucena} settings.

The above theoretical studies on atomic DB solitons have been performed in the ideal case of zero
temperature: in fact, finite-temperature induced dissipation of matter-wave solitons have basically been
studied, so far, in the simpler case of dark solitons in single-component BECs \cite{shl1,shl2,us,ft,gk,ashton}.
In particular, this problem was first addressed in Ref.~\cite{shl1} (see also Ref.~\cite{shl2}), where a
kinetic-equation approach, together with a study of the Bogoliubov-de Gennes (BdG) equations, was used.
In that work, it was found that the dark soliton center obeys an equation of motion of a harmonic oscillator,
which incorporates an {\it anti-damping} term accounting for the finite temperature effect. The presence of this
term alters the soliton trajectories so that the experimentally observed dark soliton dynamics can be
qualitatively understood: solitons either decay fast at the rims of the BEC (for high temperatures)
\cite{han1,nist,han2} or perform oscillations of growing amplitude (for low temperature)
\cite{hamburg,kip,hambcol,andreas} and eventually decay. A similar equation of motion for the dark soliton
center was also derived in Ref.~\cite{us} by applying the Hamiltonian approach of the perturbation theory
for dark matter-wave solitons \cite{djf} to the so-called {\it dissipative} Gross-Pitaevskii equation (DGPE).
This model incorporates a damping term (accounting for finite temperature), first introduced
phenomenologically by Pitaevskii \cite{lp}, and later shown to be relevant from a microscopic perspective
(see, e.g., the review \cite{npprev}). It is important to note that, as shown in Ref.~\cite{us}, the
analytical results obtained in the framework of the DGPE were found to be in very good agreement with
numerical results obtained in the framework of the stochastic Gross-Pitaevskii equation (SGPE); see, e.g.,
Ref.~\cite{stoch} for a review on the SGPE model. It should also be mentioned that while the above works
chiefly considered finite temperature effects for the case of a single dark soliton, the DGPE model and the anti-damping-incorporating ordinary differential equations (ODEs) for the soliton center were also examined in the case of multiple dark solitons. In particular, the cases of two and three oscillating and interacting, anti-damped dark solitons were considered in Ref.~\cite{dcdss}.

In the present work, we study finite-temperature dynamics of DB solitons in harmonically confined Bose gases.
In particular, we adopt an effective mean-field description and analyze theoretically and numerically a
system of two coupled DGPEs, describing the evolution of a binary quasi-one-dimensional (1D) BEC at finite
temperature. We extend the considerations of Ref.~\cite{us} and develop a Hamiltonian perturbation theory for
the two-component system at hand. This way, we obtain an equation of motion for the DB soliton center,
similar to the one derived in Refs.~\cite{shl1,us}. This equation, which includes an anti-damping term accounting
for finite temperature, provides a characteristic eigenvalue pair (i.e., a pair of solutions of the
characteristic equation associated with the linear equation of motion),
which is connected to the eigenvalue associated with the anomalous mode of the DB soliton.
Performing a Bogoliubov-de Gennes (BdG) analysis,
we show that the anomalous mode eigenvalue becomes complex as the dissipation (temperature-dependent)
parameter is introduced, leading to an instability of the DB soliton pair. The temperature-dependence
of the eigenvalues
(determined analytically) is found to be in good agreement with the one of the anomalous mode eigenvalue
(determined numerically).

Furthermore, these considerations
are generalized in the case of a DB soliton ``molecule'', composed by two-DB-solitons.
In the latter setting, both configurations featuring in-phase and
out-of-phase bright components can be obtained in the trap~\cite{ourrecent}. We illustrate their dynamical instabilities as a function of temperature and capture them
analytically by means of coupled nonlinear ODEs
accounting for the three ingredients (trap restoring force,
interaction between DB solitons and thermally induced anti-damping). We show that, due to finite temperature, the nature of their interaction (and collisions) changes: for short times individual solitons behave as repelling particles, while for longer times they gain kinetic energy and completely overlap at the collision point.
Our analytical considerations and numerical results reveal a fundamental effect: the partial stabilization that the bright (``filling'') soliton component offers to the corresponding ``bare'' dark soliton against temperature-induced anti-damping. This way, a significantly longer lifetime of the symbiotic (dark-bright) structure can be achieved, in comparison to its bare dark soliton counterpart.

The paper is structured as follows. In section II we present the model and study some of its basic properties such as
the evolution to the equilibrium state. In section III we develop the perturbation theory to derive and solve
the equation of motion for the single DB soliton; we also compare our analytical findings
to numerical results. In section IV we generalize relevant considerations to the
case of multiple DB solitons, and in section V we present our conclusions.

\section{The model and its basic properties}

\subsection{The system of dissipative Gross-Pitaevskii equations}

We consider a two-component elongated (along the $x$-direction) repulsive Bose gas, composed of two different
hyperfine states of the same alkali isotope, and confined in a highly anisotropic trap (such that the longitudinal and transverse trapping frequencies are $\omega_x \ll \omega_{\perp}$). In such a case, the system can be considered as quasi-1D and, hence, the coupling constants take their effectively 1D form, namely
$g_{jk}=2\hbar\omega_{\perp} a_{jk}$, where $a_{jk}$ denote the three $s$-wave scattering lengths (note that $a_{12}=a_{21}$) which account for collisions between atoms belonging to the same ($a_{jj}$) or different ($a_{jk}, j \ne k$) species. Let us now focus on the experimentally relevant case of a two-component BEC consisting of two different hyperfine states of $^{87}$Rb, such as the states $|1,-1\rangle$ and $|2,1\rangle$ used in the experiment of Ref.~\cite{mertes}, or the states $|1,-1\rangle$ and $|2,-2\rangle$ used in the experiments of Refs.~\cite{pea,peb}. In the first case, the scattering lengths take the values $a_{11}=100.4a_0$, $a_{12}=97.66a_0$ and $a_{22}=95.00a_0$, while in the second case the respective values are $a_{11}=100.4a_0$, $a_{12}=98.98a_0$ and $a_{22}=98.98a_0$ (where
$a_0$ is the Bohr radius). In either case, it is clear that the scattering lengths and, accordingly, the effectively 1D coupling constants take approximately the same values, say $a_{ij} \approx a$ and $g_{ij} \approx g =2\hbar\omega_{\perp} a$, respectively, which is what we will assume henceforth.

We now consider the case where the two-component Bose gas under consideration is at finite temperature. In particular, we assume that the thermal modes of energies $> \hbar \omega_{\perp}$ are at equilibrium, accounting for a heat bath in contact with the axial part of the gas, while the modes in the $x$-direction are highly occupied so that the classical field approximation is valid \cite{st1,st2}. Then, extending considerations pertinent to single-component Bose gases \cite{npprev,stoch,st1,st2} to the two-component case, we may use the following set of two coupled 1D SGPEs to describe the axial modes of the system:
\begin{eqnarray}
\!\!\!\!\!\!\!\!\!\!\!\!\!\!\!
i \hbar \partial_t \psi_j =
[1-\gamma_j (x,t)] \left( -\frac{\hbar^2}{2m} \partial_{x}^2 \psi_j +V(x) -\mu_j
+ g \sum_{k=1}^2 |\psi_k|^2\right) \psi_j
+ \eta_j(x,t).
\label{sgpes}
\end{eqnarray}
Here, $\psi_j(x,t)$ ($j=1,2$) are complex order parameters characterizing each component of the binary Bose gas, $m$ is the atomic mass, $\mu_j$ are the chemical potentials, while $V(x)=(1/2)m\omega_x^2 x^2$ is the external
trapping potential. Furthermore, $\eta_j(z,t)$ are complex Gaussian noise terms with correlations of the form
$\langle \eta_j^{\ast}(x,t)\eta_j(x',t') \rangle = 2\hbar \gamma_j(x,t)k_B T \delta(x-x')\delta(t-t')$, where brackets denote averaging over different realizations of the noise. The strength of the latter can be calculated {\it ab initio} by the Keldysh self-energy \cite{st1}; for thermal clouds close to equilibrium, the relevant integrals determining the dissipation $\gamma_j(x,t)$ can be expressed as follows:
\begin{eqnarray}
\!\!\!\!\!\!\!\!\!\!\!\!\!\!\!
\gamma_j(x)&=&\pi^2 \beta g^2 \int \frac{dk_1}{2\pi} \int \frac{dk_2}{2\pi} \int \frac{dk_3}{2\pi}
2\pi \delta(k_1-k_2-k_3)\delta(\epsilon_c^{(j)}+\epsilon_1^{(j)}-\epsilon_2^{(j)}-\epsilon_3^{(j)})
\nonumber \\
&\times & [N_1(1+N_2)(1+N_3)+(1+N_1)N_2N_3],
\label{gamma_int}
\end{eqnarray}
where $\beta = 1/k_B T$, $\epsilon_c^{(j)}$ are the condensate energies, $\epsilon^{(j)}_{n}$ are the energies of the $n$-th excited states,
$N^{(j)}_{n}=[\exp(\beta(E^{(j)}_{n}+V(x)+2g\sum_{k=1}^2 \langle |\psi_k|^2 \rangle -\mu_j))-1]^{-1}$
are Bose-Einstein distributions, while $E^{(j)}_{n}$ and $k_{n}=\sqrt{2 m E_{n}/\hbar^2}$ denote, respectively, the kinetic energies and momenta of single particles in the $n$-th excited state. Physically speaking, Eq.~(\ref{gamma_int}) describes the exchange of atoms between the thermal clouds and the condensates due to elastic collisions; notice that in the above description we have taken into regard exchanges up to the third excited state while, to leading order approximation, we have omitted exchanges between the different hyperfine states (in other words, we have considered the simplest situation where each condensate component interacts with its own thermal cloud).

Under the above assumptions, the dissipation terms $\gamma_j(x)$ may in principle be calculated numerically, for several temperatures, as was done in the case of a single-component Bose gas in Refs.~\cite{us}. In this work, it was shown that, sufficiently close to the trap center (i.e., in the interval $[-R/2,~R/2]$, where $R$ is the Thomas-Fermi radius),  the dissipation takes approximately constant values for a relatively wide range of temperatures. Furthermore, as shown in Refs.~\cite{shl1,shl2,gk} (see also the discussion in Ref.~\cite{us,gk} and, more recently, in Ref.~\cite{ashton}), the value of $\gamma$---which determines the dark soliton's life time---scales with temperature as $\gamma \propto T^\alpha$,
with $1 < \alpha < 4$; note that the case $\gamma \propto T^4$ corresponds to the regime $k_B T \ll \mu$, while the case
$\gamma \propto T$ corresponds to the regime $k_B T \gg \mu$ (where $\mu$ is the chemical potential of the background Bose liquid).

Taking into regard the above findings, below we will consider the situation where both dissipative terms $\gamma_j$ are constant: such an assumption is consistent with our scope, i.e., to analyze the dynamics of the DB-soliton near the center of the trap. Furthermore, based on the fact that simulations investigating soliton dynamics in the framework of the SGPE model were found to be in fairly good agreement with analytical and numerical results relying on the respective DGPE model, below we will omit the noise terms $\eta_j(x,t)$; this way, we will use the following system of two coupled DGPEs to describe the DB soliton dynamics in the two-component Bose gas at finite temperatures:
\begin{eqnarray}
(i-\gamma_j) \hbar \partial_t \psi_j =
\left( -\frac{\hbar^2}{2m} \partial_{x}^2 \psi_j +V(x) -\mu_j + g \sum_{k=1}^2 |\psi_k|^2\right) \psi_j.
\label{model}
\end{eqnarray}
%
%
Note that the above model was recently used in Ref.~\cite{jap}, where the quantum Kelvin-Helmholtz instability of a two-component BEC was studied.

The system of Eqs.~(\ref{model}) can be expressed in dimensionless form as follows. Measuring the densities $|\psi_j|^2$, length, time and energy in units of $2a$, $a_{\perp} = \sqrt{\hbar/\omega_{\perp}}$, $\omega_{\perp}^{-1}$ and $\hbar\omega_{\perp}$, respectively, Eqs.~(\ref{model}) become:
\begin{eqnarray}
(i-\gamma_d) \partial_t u_d  &=& -\frac{1}{2} \partial_{x}^2u_d  + V(x)u_d
+(|u_d|^2 + |u_b|^2 -\mu) u_d,
\label{deq1}
\\
(i-\gamma_b) \partial_t u_b  &=& -\frac{1}{2} \partial_{x}^2u_b +V(x)u_b
+ (|u_b|^2 + |u_d|^2- \mu-\Delta) u_b,
\label{deq2}
\end{eqnarray}
where we have used the notation $\psi_1 = u_d$ and $\psi_2 = u_b$, indicating that the component $1$ ($2$) is
supposed to support a dark (bright) soliton, and the respective chemical potentials are now $\mu_1=\mu_d=\mu$ and $\mu_2=\mu_b=\mu+\Delta$; in our considerations below we assume that $\mu_d>\mu_b$, i.e., $\Delta =-|\Delta|<0$. Finally, the external potential in Eqs.~(\ref{deq1})-(\ref{deq2}) takes the form $V(x)=(1/2)\Omega^{2} x^{2}$, where
$\Omega = \omega_x/\omega_\perp \ll 1$ is the normalized trap strength; the latter, along with the thermally
induced damping parameters $\gamma_{d,b}$, are considered to be small parameters of the system (these will be treated
as formal perturbation parameters in our analytical approximation -- see below).

We should add a comment here about the relevant range of values of the parameter $\gamma$. A number of recent
experiments, including the ones in Hamburg~\cite{hamburg,hambcol}, Heidelberg~\cite{kip,andreas} and Pullman~\cite{pea,peb}, have focused on regimes of very low temperature where the effect of the term associated with $\gamma$ is imperceptible (over the experimentally relevant time scales). The focus of these experiments was on the soliton dynamics and an effort was made (by operating at $T/T_c \leq 0.1$) to correspondingly minimize the thermal effects. It is easier to appreciate the latter features in the context of the earlier experiments of the Hannover group \cite{han1,han2}, which were conducted in the regime of $T/T_c \approx 0.5$. In that realm, the relevant values of $\gamma$ can be estimated to be up to $10^{-2}$~\cite{cockburn}. In what follows, we will treat $\gamma$ generally as a free parameter, in order to illustrate the available wealth of bifurcation and dynamical phenomena of this system. Nevertheless, the reader more keen on the physical applications of the model to the physics of finite-temperature BECs should keep in mind the above values as a guideline towards the parameter regimes pertinent therein. We finally note that our analysis may also be used as a theoretical basis for understanding results of future experiments on dark and dark-bright solitons exploring finite-temperature effects (see, e.g., discussion in the Supplemental Material of Ref.~\cite{pea}).

\subsection{Relaxation to the ground state of the system}

Since our purpose is to study the dissipative dynamics of DB solitons in this setting, it is
natural to consider at first the dynamics of the pertinent background wave functions, namely a Thomas-Fermi
(TF) wave function for the $u_d$ component and a zero wave function for the $u_b$ component.
In particular, we will show that the coupled DGPEs Eqs.~(\ref{deq1})-(\ref{deq2}), similarly to
their one-component counterpart (see, e.g., discussion in Ref.~\cite{bur}), describe a {\it relaxation process}.
Namely, as a result of the finite temperature, the two components,
starting (at $t=0$) from suitable initial conditions,
will evolve so that, at sufficiently large times,
$u_d$ will converge towards a TF cloud with the prescribed value of the
chemical potential $\mu$, while $u_b$ will
vanish.

To show that this is the case indeed, we examine the peak amplitudes
$U_{d,b}(t)$ of the wave functions $u_{d,b}(x=0,t)$, corresponding to their (absolute) values at the center
of the trap (i.e., at $x=0$, where $V(x)=0$ as well), and
assume respective phases $\theta_{d,b}(t)$.
The evolution equations for $U_{d,b}(t)$ and $\theta_{d,b}(t)$, which can directly be obtained by
introducing the ansatz $u_{d,b} = U_{d,b}(t)\exp[-i\theta_{d,b}(t)]$ into
Eqs.~(\ref{deq1})-(\ref{deq2}), are of the form:
\begin{eqnarray}
&&\dot{U}_{d,b} + \gamma_{d,b} U_{d,b} \dot{\theta}_{d,b} = 0,
\label{db1} \\
&&\gamma_d \dot{U}_{d} - \dot{\theta}_d U_d + (U_d^2+U_b^2 -\mu)U_d  =0
\label{db2} \\
&&\gamma_b \dot{U}_{b} - \dot{\theta}_b U_b + (U_b^2+U_d^2 -\mu-\Delta) U_b =0,
\label{db3}
\end{eqnarray}
where overdots denote time derivatives. Next, utilizing Eqs.~(\ref{db1}), we obtain from
Eqs.~(\ref{db2})-(\ref{db3}) the following system:
\begin{eqnarray}
\dot{U}_{d}&=&-\tilde{\gamma}_d \left(U_{d}^2+U_{b}^2-\mu \right)U_{d},
\label{ud0}
\\
\dot{U}_{b}&=&- \tilde{\gamma}_b \left(U_{d}^2+U_{b}^2-\mu -\Delta \right)U_{b},
\label{ub0}
\end{eqnarray}
where $\tilde{\gamma}_{d,b}\equiv \gamma_{d,b}/(1+\gamma_{d,b}^2)$. It is clear that that the system of
Eqs.~(\ref{ud0})-(\ref{ub0}) has a fixed point
$(U_{d0},~U_{b0})=(\sqrt{\mu},~0)$ [a similar analysis can be done for
the fixed point $(U_{d0},~U_{b0})=(0,\sqrt{\mu+\Delta})$]. The evolution of small
perturbations $U_{d1,b1}$ around this fixed point can then readily be found introducing the ansatz
$U_{d0}(t)=\sqrt{\mu}+U_{d1}(t)$ and $U_{b0}=U_{b1}(t)$ into Eqs.~(\ref{ud0})-(\ref{ub0}) and linearizing
with respect to $U_{d1,b1}$; this way, we can easily solve the equations for $U_{d1,b1}$ and finally
obtain the following approximate expressions for the peak amplitudes of the wave functions:
\begin{eqnarray}
U_{d}(t)& \approx &\sqrt{\mu}+(U_{d}(0)-\sqrt{\mu}){\rm e}^{-2\tilde{\gamma}_{d} \mu t},
\label{ud00}
\\
U_{b}(t)& \approx &U_{b}(0){\rm e}^{-\tilde{\gamma}_{b} |\Delta| t},
\label{ub00}
\end{eqnarray}
where $U_{d,b}(0)$ are initial conditions. Thus, at sufficiently large times,
the peak amplitude of $u_d$ will decay to the value
$\sqrt{\mu}$, while the one of $u_b$ will become zero. Accordingly, during the
relaxation to equilibrium
process, one may expect the following type of evolution towards relaxation.
If the $u_d$ component is initially a
Thomas-Fermi (TF) cloud of amplitude $U_d(0)$, its density will evolve as,
\begin{eqnarray}
|u_{d}(x,t)|^2 \approx U_d^2(t) - V(x).
\label{udf}
\end{eqnarray}
On the other hand, if the $u_b$ component has initially the form of an
arbitrary localized function,
e.g., a Gaussian, of amplitude $U_b(0)$, it will asymptotically
approach the trivial stationary state.


\begin{figure}[tbp]
\begin{center}
\includegraphics[scale=0.45]{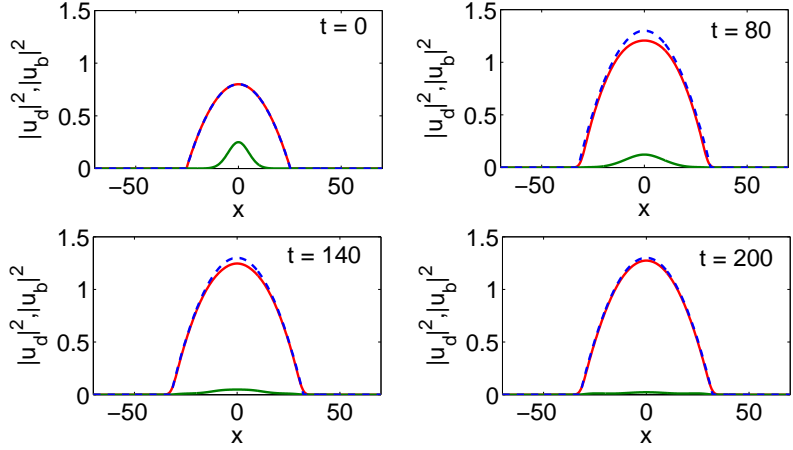}
\caption{(Color online) Time evolution of a state characterized by the densities
$u_d(x,0)|^2=U_d^2(0)-(1/2)\Omega^2 x^2$ and $|u_b(x,0)|^2=U_b^2(0) \exp\left[-2(x/w)^2\right]$,
with parameter values $U_d(0)=0.86$, $U_b(0)=0.6$, $\Omega=0.05$ and $w=10$. The solid lines show
the density of the dark (red) and bright (green) component, while the dashed line shows the
analytical result of Eq.~(\ref{ud00}). The other parameter values used in Eqs.~(\ref{deq1})-(\ref{deq2}) are
$\mu=1.3$, $|\Delta|=0.1$ and $\gamma_d=\gamma_b=0.05$. }
\label{relax}
\end{center}
\end{figure}


The above predictions can be directly compared to numerical simulations. In particular, in Fig.~\ref{relax} we
show the evolution of a state characterized by the initial densities $|u_d(x,0)|^2=U_d^2(0)-(1/2)\Omega^2 x^2$
and $|u_b(x,0)|^2=U_b^2(0) \exp\left[-2(x/d)^2\right]$, with parameter values $U_d(0)=0.86$, $U_b(0)=0.6$, $\Omega=0.05$ and $d=10$;
as found by direct numerical integration of Eqs.~(\ref{deq1})-(\ref{deq2}), with $\mu=1.3$, $|\Delta|=0.1$ and
$\gamma_d=\gamma_b=0.05$. The figure clearly shows the validity of our analytical approximations:
the $u_d$ component develops into a TF cloud with chemical potential
$\mu=1.3$, with the numerically found
density profile [solid (red) line] being in fairly good agreement with the analytical prediction of Eq.~(\ref{udf})
(dashed line); on the other hand, $u_b$-component [solid (green) line] vanishes at $t\approx 200$, a time consistent
with the slow time scale $t_{\ast} \equiv (\tilde{\gamma}_{b} |\Delta|)^{-1} \approx 200$ suggested by Eq.~(\ref{ub00}).

\section{Dissipative Dynamics of a Single Dark-Bright Soliton}

\subsection{Analytical results}

Having studied the relaxation process described by Eqs.~(\ref{deq1})-(\ref{deq2}), we
will now proceed to investigate, in the same framework, the dissipative dynamics of DB solitons.
We will assume that the dark soliton is
on top of an already formed TF cloud with the equilibrium
density $|u_{d,TF}|^2 = \mu-V(x)$; this way, the
density $|u_d|^2$ in Eqs.~(\ref{deq1})-(\ref{deq2}) is substituted by $|u_d|^2 \rightarrow |u_{d,TF}|^2 |u_d|^2$.
Furthermore, we introduce the transformations $t \rightarrow \mu t$, $x \rightarrow {\sqrt{\mu}}x$,
$|u_b|^2 \rightarrow \mu^{-1} |u_b|^2$, and cast Eqs.~(\ref{deq1})-(\ref{deq2}) into the following form:
%
%
\begin{eqnarray}
i\partial_tu_d+\frac{1}{2}\partial^2_xu_d-\left(|u_d|^2+|u_b|^2-1\right)u_d &=& R_d,
\label{u_d}
\\
i\partial_tu_b+\frac{1}{2}\partial^2_xu_b-\left(|u_b|^2+|u_d|^2-\tilde{\mu}\right)u_b &=& R_b,
\label{u_b}
\end{eqnarray}
where $\tilde{\mu}=1+\Delta/\mu$ and
\begin{eqnarray}
R_d &\equiv& (2\mu^2)^{-1}[2(1-|u_d|^2)V(x)u_d+V'(x)\partial_xu_d]+\gamma_d\mu^{-1}\partial_t u_d,
\label{R_d}
\\
R_b &\equiv& \mu^{-2}[(1-|u_d|^2)V(x)u_b+\mu\gamma_b\partial_t u_b].
\label{R_b}
\end{eqnarray}
while $V'(x)\equiv dV/dx$. Equations~(\ref{u_d})-(\ref{u_b}) can be viewed as
a system of two coupled perturbed nonlinear Schrodinger (NLS) equations, with
perturbations given by Eqs.~(\ref{R_d})-(\ref{R_b}). In the absence of the
perturbations, i.e., at zero temperature ($\gamma_b=\gamma_d=0$) and for the
homogeneous system ($V(x)=0$) subject to the boundary conditions
$|u_d|^2\rightarrow1$ and $|u_b|^2\rightarrow0$ as $|x|\rightarrow\infty$,
the NLS Eqs.~(\ref{u_d})-(\ref{u_b}) possess an exact analytical one-DB-soliton
solution of the following form:
\begin{eqnarray}
u_d(x,t) &=& \cos\phi\rm tanh[D(x-x_0(t))]+i\sin\phi,
\label{db_u_d}
\\
u_b(x,t) &=& \eta\rm sech[D(x-x_0(t))]exp[ikx+i\theta(t)],
\label{db_u_b}
\end{eqnarray}
where $\phi$ is the dark soliton's phase angle, $\cos\phi$ and $\eta$ represent the amplitudes of the dark and bright solitons, and $D$ and $x_0(t)$ are associated with the inverse width and the center position of the DB soliton.
Furthermore, $k=D\rm tan\phi=const$ and $\theta(t)$ are the wavenumber and
phase of the bright soliton, respectively. The above parameters of the
DB-soliton are connected through the following equations:
\begin{eqnarray}
D^2 &=& \cos^2\phi-\eta^2,
\label{D}
\\
\dot{x_0} &=& D\rm tan\phi,
\label{d_x_0}
\\
\theta(t) &=& \frac{1}{2}(D^2-k^2)t+(\Delta/\mu)t,
\label{theta}
\end{eqnarray}
with $\dot{x}_0$ denoting the DB soliton velocity. Notice that the
amplitude $\eta$ of the bright soliton, the chemical potential $\mu$ of the
dark soliton, as well as the (inverse) width parameter $D$ of the DB soliton
are connected to the number of atoms of the bright soliton by means of the
following equation:
\begin{eqnarray}
N_b&\equiv&\int_{\mathbb{R}}|u_b|^2dx=\frac{2\sqrt{\mu}\eta^2}{D}.
\label{nb}
\end{eqnarray}

Let us now employ the Hamiltonian approach
of the perturbation theory for the matter-wave solitons to study the dissipative dynamics of DB solitons. We start by considering the Hamiltonian (total energy) of the system of Eqs.~(\ref{u_d})-(\ref{u_b}), in the
absence of the perturbations (i.e., for $R_b=R_d=0$), namely,
\begin{eqnarray}
E &=& \frac{1}{2}\int_{-\infty}^{+\infty} \mathcal{E} dx, \nonumber \\
\mathcal{E} &=& |\partial_{x} u_d|^2+|\partial_{x} u_b|^2+(|u_d|^2+|u_b|^2-1)^2
- 2(\tilde{\mu}-1)|u_b|^2.
\label{energy}
\end{eqnarray}
The energy of the system, when calculated for the DB soliton solution of Eqs.~(\ref{db_u_d})-(\ref{db_u_b}), takes the following form:
\begin{eqnarray}
E = \frac{4}{3}D^3+\chi \left(\frac{1}{2}D^2\rm sec^2\phi-\frac{\Delta}{\mu}\right), \qquad
\chi = \frac{N_b}{\sqrt{\mu}}.
\label{E}
\end{eqnarray}
We now consider an adiabatic evolution of the DB soliton and, particularly, we assume that, in the presence of the perturbations of Eqs.~(\ref{R_d})-(\ref{R_b}), the DB soliton parameters become slowly-varying unknown functions of time $t$. Thus, the DB soliton parameters become $\phi\rightarrow\phi(t)$, $D\rightarrow D(t)$ and, as a result, Eqs.~(\ref{D})-(\ref{d_x_0}) read:
\begin{eqnarray}
D^2(t) &=& \cos^2\phi(t)-\frac{\chi}{2}D(t),
\label{d(t)}
\\
\dot{x}_0(t) &=& D(t)\rm tan\phi(t),
\label{x_0(t)}
\end{eqnarray}
where we have used Eq.~(\ref{nb}). The evolution of the parameters $\phi(t)$, $D(t)$ and $x_0(t)$ can be found by means of the evolution of the DB soliton energy. In particular, employing Eq.~(\ref{E}), it is readily found that
\begin{eqnarray}
\frac{dE}{dt} &=& 4\dot{D}D^2+
\chi D \sec^2\phi(\dot{D}+D\dot{\phi}\rm tan\phi).
\label{dE_dt}
\end{eqnarray}
On the other hand, using Eqs.~(\ref{u_d})-(\ref{u_b}) and their complex conjugates, it can be found that the evolution of the DB soliton energy, due to the presence of the perturbations, is given by:
\begin{eqnarray}
\frac{dE}{dt} &=& -2{\rm Re}\left\{\int_{\mathbb{R}}(R_d^{*}\partial_tu_d+R_b^{*}\partial_tu_b)dx \right\},
\label{dotE}
\end{eqnarray}
where the asterisk denotes complex conjugation. Substituting $R_d$ and $R_b$ into Eq.~(\ref{dotE}) and evaluating the integrals, we finally obtain from Eqs.~(\ref{dE_dt})-(\ref{dotE}) the following result:
\begin{eqnarray}
4\dot{D}D^2+
\chi D\sec^2\phi(\dot{D}+D\dot{\phi}\rm tan\phi) =
\nonumber\\
\frac{1}{\mu^2}\left(2\cos^3\phi\sin\phi-
\chi D\sin\phi\cos\phi\right)V'(x_0)
\nonumber\\
-\frac{8}{3}\frac{\gamma_d}{\mu}D^3\sin^2\phi-\frac{2}{3}\frac{\gamma_b}{\mu}
\chi D^4\rm tan^2\phi. &&
\label{perturbed}
\end{eqnarray}
Equation~(\ref{perturbed}), together with Eqs.~(\ref{d(t)})-(\ref{x_0(t)}), constitute a system of equations for the unknown soliton parameters $\phi(t)$, $D(t)$ and $x_0(t)$.
In the case of a DB soliton near the center of the trap with an almost ``black" dark-soliton-component (i.e., $x_0\approx0$ and $\cos\phi\approx1$), the above system has a fixed point $x_{0,eq}=0$ and $\phi_{eq}=0$, and
\begin{eqnarray}
D_{eq} &=& \sqrt{1+\left(\frac{\chi}{4}\right)^2}-\frac{\chi}{4}.
\end{eqnarray}
Considering now small perturbations around the fixed points,
i.e., $x_0 \rightarrow 0+x_0$, $\phi \rightarrow 0+\phi$ and $D \rightarrow D_{eq}+D_1$,
we linearize Eqs.~(\ref{d(t)})-(\ref{x_0(t)}) and Eq.~(\ref{perturbed}) with respect to $x_0$, $\phi$ and $D_1$, and obtain the following results:
\begin{eqnarray}
D_1 &=& -\tilde{D}\phi^2, \qquad \tilde{D} \equiv \left(2D_{eq}+\frac{\chi}{2}\right)^{-1},
\label{D_1}
\\
\dot{\phi} &=& \frac{-2+
\chi D_{eq}}{D_{eq}[-8D_{eq}\tilde{D}-
\chi (2\tilde{D}-D_{eq})]}V'(x_0)
\nonumber \\
&+& \frac{\frac{2}{3\mu}D_{eq}^3\left(4\gamma_d+
\chi \gamma_b D_{eq}\right)\phi}{D_{eq}[-8D_{eq}\tilde{D}-
\chi (2\tilde{D}-D_{eq})]},
\label{phi_1}
\end{eqnarray}
\begin{eqnarray}
\dot{x}_0 &=& D_{eq}\phi.
\label{dotx0}
\end{eqnarray}
Differentiating Eq.~(\ref{dotx0}) with respect to time once, and
using Eqs.~(\ref{phi_1})-(\ref{dotx0}), we obtain after some straightforward
algebraic manipulations the following equation of the motion for the DB
soliton center $x_0$:
\begin{eqnarray}
\ddot{x}_0-a\dot{x}_0+\omega_{\rm osc}^2x_0 &=& 0,
\label{ddotx0}
\end{eqnarray}
where the oscillation frequency $\omega_{\rm osc}$ and the anti-damping parameter $a$ are respectively given by:
\begin{eqnarray}
\omega_{\rm osc}^2 = \Omega^2 \left(\frac{1}{2}- \frac{\chi}{\chi_0}
\right), \qquad \chi_0 \equiv 8 \sqrt{1+\left( \frac{\chi}{4}\right)^2},
\label{omega}
\end{eqnarray}

\begin{eqnarray}
a = \frac{2}{3}\mu\left(\gamma_d-
\frac{1}{8}\chi^2 \gamma_b\right) +\frac{4}{3} \frac{\chi}{\chi_0} \mu\left(\gamma_b-\gamma_d+
\frac{1}{8}\chi^2 \gamma_b\right).
\label{aa}
\end{eqnarray}
%
%
%
We note that in the absence of dissipation [$a=0$ in Eq.~(\ref{ddotx0})], Eq.~(\ref{omega}) recovers the results of Ref.~\cite{BA01}: according to this work, if both components are confined in the same harmonic trap of strength $\Omega$ then a DB-soliton oscillates around the trap center with the frequency $\omega_{\rm osc}$, given in Eq.~(\ref{omega}). 

It is clear that the nature of the soliton trajectories $x_0(t)$ as predicted by Eq.~(\ref{ddotx0}) depend on whether the roots of the auxiliary equation $s^2 - a s + \omega_{\rm osc}^2 = 0$ are real or complex. The roots are given by
\begin{equation}
s_{1, 2} = \frac{1}{2} \left( a \pm \sqrt{a^2-a_{\rm cr}^2} \right),
\,\,\,\,\,
a_{\rm cr} \equiv 2\omega_{\rm osc},
\label{s12}
\end{equation}
with the discriminant $\mathcal{D} \equiv a^2-a_{\rm cr}^2$ determining the type of the motion.
In particular, we identify different temperature-dependent damping
regimes: the subcritical weak anti-damping regime ($\mathcal{D} <0$,
$a<a_{cr}$), the critical regime ($\mathcal{D}=0$, $a=a_{cr}$), and
the super-critical strong antidamping  regime ($\mathcal{D} >0$,
$a>a_{cr}$). In the first regime the soliton performs oscillations of
growing amplitude, with $x_0(t)\propto \exp(at)\cos(\omega_{\rm
  osc}t)$, while in the latter two regimes the soliton follows an
exponentially growing trajectory, i.e., $x_0(t)\propto \exp(s_{1,2}
t)$ (with $s_{1,2} \in \mathbb{R}$), and decays at the rims of the
condensate cloud (see also below).


\subsection{Numerical results}

We now turn to a numerical examination of the above findings. First, we will show that our analytical predictions are supported by a linear stability analysis around the stationary DB soliton, say ${\bf u_0} \equiv ({\rm u,~v})^T$
[see Eqs.~(\ref{db_u_d})-(\ref{db_u_b}) for $\phi=0$ and $x_0=0$]. For such a state, the right-hand side of
Eqs.~(\ref{deq1})-(\ref{deq2}) still vanishes and, thus, stationary DB solitons are exact solutions
of the problem with $\gamma_{d,b} \ne 0$. We obtain this solution by means of a fixed point algorithm and then find
the linearization spectrum around the stationary DB soliton state as follows. We introduce
the ansatz
\begin{equation}
{\bf u} ={\bf u_0} + \epsilon [\exp(\lambda t) {\bf a}(x) + \exp(\lambda^{\ast} t) {\bf b}^{\ast}(x)],
\label{b}
\end{equation}
into the DGPEs Eqs.~(\ref{deq1})-(\ref{deq2}) (here $\{\lambda,({\bf a},~{\bf b})\}$ define an eigenvalue-eigenvector pair, and $\epsilon$ is a formal small parameter), and then solve the ensuing BdG eigenvalue problem.

\begin{figure}
\begin{center}
\includegraphics[scale=0.5]
{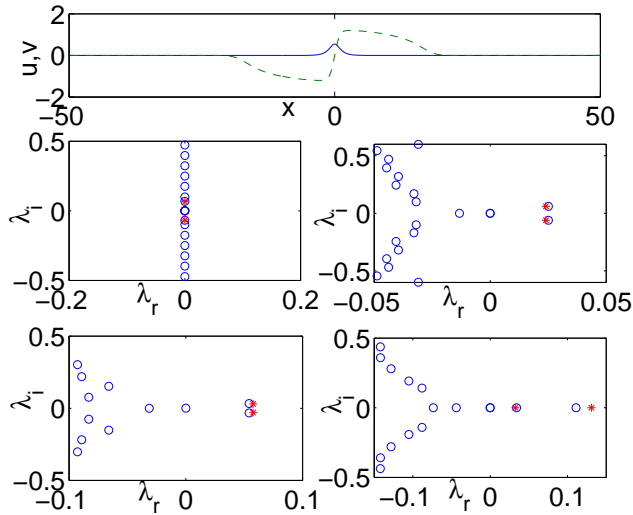}
\caption{(Color online)
The top panel depicts the stationary solution for a single DB-soliton
for $\mu=1.5$, $|\Delta|=0.6$ and $\Omega=0.1$. The dark (bright) components
are shown by the dashed green (solid blue) lines.
The middle and bottom panels are four spectral planes, corresponding to different
values of $\gamma=\gamma_d=\gamma_b$, for the single dark-bright soliton
stationary states: in the
middle left panel $\gamma=0$ (the zero-temperature Hamiltonian case), in
the middle right $\gamma=0.05$, in the bottom left $\gamma=0.12$, and in
the bottom right $\gamma=0.17$. The (red) stars highlight the anomalous
mode (of the Hamiltonian case) eigenvalues.}
\label{fig1}
\end{center}
\end{figure}

In Fig.~\ref{fig1}, we observe a prototypical realization of a stationary DB soliton
in a trap of strength $\Omega=0.1$ (for simplicity, we consider the case with $\gamma_d=\gamma_b= \gamma $).
Notice that upon the variations of $\gamma$ (and hence of temperature) considered in the figure, the solution profile
does not change, as mentioned above; however, the linearization problem and its eigenvalues significantly depend on the value of $\gamma$, as is shown in the four bottom panels
of Fig.~\ref{fig1}. In the zero-temperature (Hamiltonian) case
of $\gamma=0$,
all eigenvalues are imaginary. Furthermore,
the oscillatory motion of
a single DB soliton
in the trap \cite{BA01} (see also recent work in Refs.~\cite{pea,peb,ourrecent}) is spectrally
associated with the existence of a single {\it anomalous} (alias {\it negative Krein sign}, ``translational'') {\it mode}
in the linearization around the stationary soliton.
In analogy to
the case of dark solitons (see e.g. Refs.~\cite{shl1,andreas,dcdss}),
this anomalous mode possesses a frequency identical to the frequency of the DB soliton oscillation,
i.e., $\omega_{\rm AM} \equiv {\rm Im}(\lambda_{\rm AM}) = \omega_{\rm osc}$.

\begin{figure}
\begin{center}
\includegraphics[scale=0.5]
{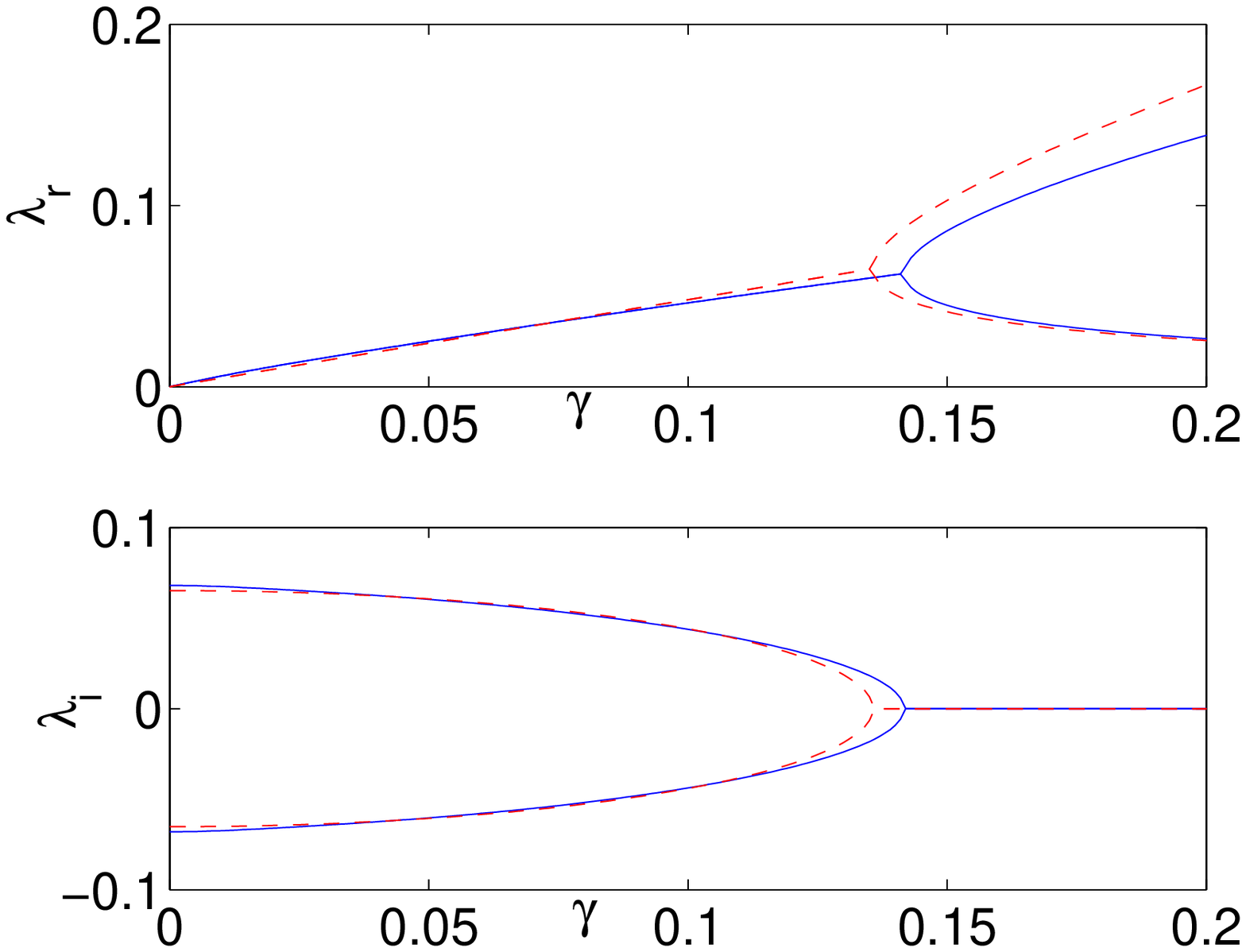}
\caption{(Color online)
The real part (top) - instability growth rate- and imaginary part (bottom) -oscillation
frequency- of the unstable eigenmode of the linearization around a stationary DB
soliton as a function of the parameter $\gamma$ (parameter values are as in Fig.~\ref{fig1}).
Solid (blue) lines indicate the full numerical result, while dashed (red) ones the analytical results of
Eq.~(\ref{s12}).
For $\gamma<\gamma_{cr} \approx 0.141$ (subcritical regime), the complex conjugate pair is responsible for soliton oscillations of growing amplitude. For $\gamma \ge \gamma_{cr}$ (critical/super-critical regimes) the collision of the complex conjugate pair of eigenvalues creates a real pair, and
the dynamics involves purely exponential growth. The parameter values are $\mu=1.5$, $|\Delta|=0.6$, and $\Omega=0.1$.
}
\label{fig2}
\end{center}
\end{figure}

Our analytical approximation for $\omega_{\rm osc}$
is tested against the numerical results for ${\rm Im}(\lambda_{\rm AM})$,
both in the
case examples of Fig.~\ref{fig1}, as well as in the parametric dependence results of Fig.~\ref{fig2}.
It is clear from the spectral plots (middle and bottom panels of Fig.~\ref{fig1}) that, as soon as $\gamma \neq 0$, the relevant anomalous eigenmode (indicated by red stars in Fig.~\ref{fig2}) becomes {\it complex}, leading to soliton oscillations of growing amplitude; this behavior, which corresponds to the ``subcritical'' regime mentioned above, is
similar to the case of dark solitons \cite{us,dcdss} and in accordance with rigorous results pertaining to dissipative
NLS systems
\cite{sand}.
If $\gamma$ is increased beyond a critical point, namely
$\gamma_{cr} \approx 0.141$,
the relevant eigenvalue pair collides
with the real axis, leading to the emergence of a pair of real eigenvalues (cf. bottom right panel of Fig.~\ref{fig1} and Fig.~\ref{fig2}).
This corresponds to the ``super-critical''
regime
mentioned above (see also Refs.~\cite{us,dcdss}),
where the divergence of the soliton from its center equilibrium is purely exponential. Notice that the analytical predictions for the relevant unstable eigenvalue (and the oscillatory or purely exponential divergence from the equilibrium position) in Figs.~\ref{fig1} and \ref{fig2} are generally fairly accurate, although their accuracy is decreasing as $\gamma$ gets larger; this can be understood by the fact that our analytical approximation relies on the smallness of
$\gamma$ which was treated as a small parameter of the problem within our perturbation theory approach.

\begin{figure}[tbp]
\begin{center}
\includegraphics[scale=0.4]{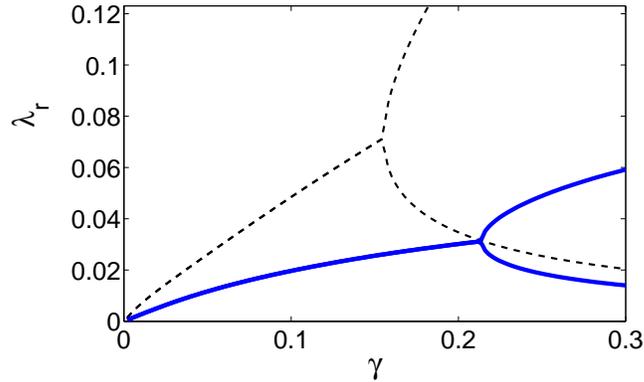}
\caption{(Color online) Comparison of the bifurcation diagrams $\lambda_r(\gamma)$ for a stationary DB soliton [solid (blue) line] and a ``bare'' dark soliton [dashed (black) line] as obtained by the BdG analysis. It is clear that the presence of the bright (``filling'') component drifts $\gamma_{cr}$ towards larger values, acquiring also smaller values of the instability growth rate $\lambda_r$ as compared to the ones found for the dark soliton. The parameter values are $\Omega=0.1$, $\mu=1.5$ and $|\Delta|=0.1$.}
\label{2b}
\end{center}
\end{figure}

Last but not least, the role of the bright-soliton component in the dynamics should be highlighted in connection to the case of a dark soliton in a single-component condensate (where the bright soliton is absent). It can be directly seen from Eq.~(\ref{aa}) that the anti-damping effect is always {\it weaker} for the DB soliton in comparison to the dark one (at least in the case $\gamma_d=\gamma_b=\gamma$ we consider herein). Hence, the lifetime of the DB soliton is always longer than that of the dark soliton and, in fact, it becomes larger, as the bright-soliton component ``filling'' of the dark one becomes stronger. This {\it partial stabilization} of the dark soliton evolution by means of its symbiotic second component is clearly illustrated in Fig.~\ref{2b}, where the bifurcation diagrams for the DB soliton are directly compared to the ones corresponding to the ``bare'' dark soliton. It is clear that the whole bifurcation diagram for the DB soliton is ``drifted'' towards larger values of $\gamma$ (e.g., $\gamma_{\rm cr}=0.212$ for the DB soliton and $\gamma_{\rm cr}=0.155$ for the dark soliton), acquiring also smaller values of the instability growth rate $\lambda_r$ as compared to the ones found for the dark soliton for the same values of the temperature parameter $\gamma$. This is a clear indication that the ``filled'' dark soliton in a two-component BEC is more robust in the presence of finite temperature than a ``bare'' dark soliton in a single-component BEC. This is one of the principal findings of the present work.

\begin{figure}
\centering\includegraphics[scale=0.5]
{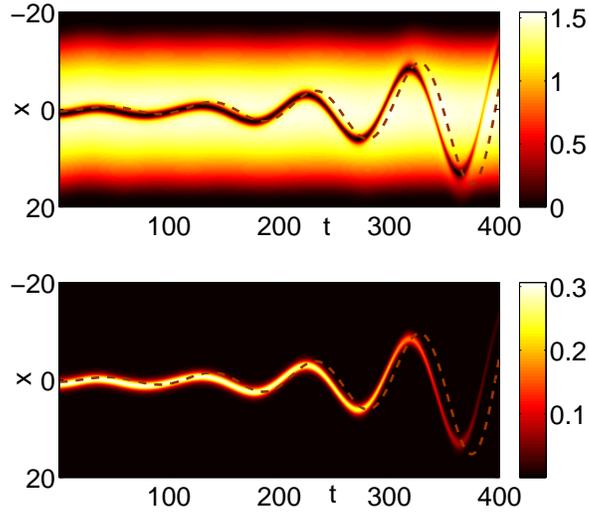}
\caption{(Color online)
Contour plot showing the space-time evolution of the density
in the single
DB soliton case. The top panel represents the dark soliton and the bottom one the bright
soliton, with $\gamma=0.02$ (subcritical regime) and $\Omega=0.1$. The soliton
is initially placed at $x_0(0)=0.4$. The dashed line represents the analytical result
of Eq.~(\ref{ddotx0}), namely
$x_0(t) = x_0(0)\exp[(a/2)t] \cos(\omega_{\rm osc} t)$.
The parameter values are $\mu=1.5$, $|\Delta|=0.6$, and $\Omega=0.1$.
}
\label{fig7}
\end{figure}

\begin{figure}
\centering\includegraphics[scale=0.5]
{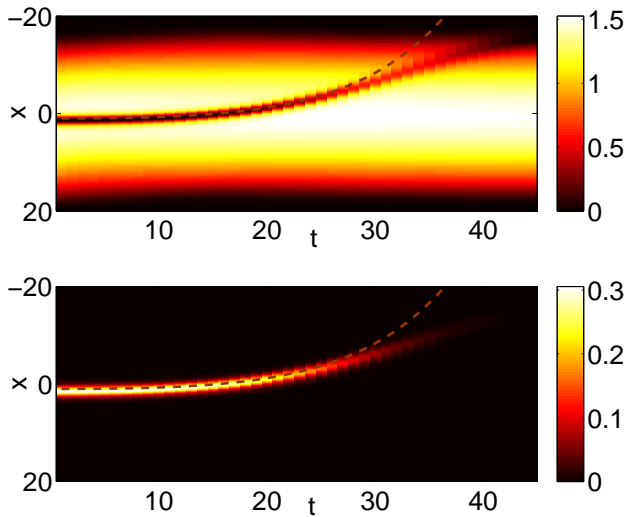}
\caption{(Color online)
%
Similar to Fig.~\ref{fig7}, but now for the super-critical case of
$\gamma=0.15$, which
shows the exponential divergence of the soliton center. The soliton is
initially placed at $x_0(0)=1$. The dashed line represents the analytical result
of Eq.~(\ref{ddotx0}), namely $x_0(t) = \frac{x_0(0)}{s_2 - s_1}\left[ s_2 \exp(s_1 t)-s_1 \exp(s_2 t)  \right]$.
}
\label{fig8}
\end{figure}

Our analytical predictions were also tested against direct numerical simulations illustrating the evolution of the single DB soliton, both for the sub-critical case of oscillatory growth (see Fig.~\ref{fig7}), and for the supercritical case of purely exponential growth (see Fig.~\ref{fig8}). In both cases it can be seen that the dashed line
corresponding to the analytical solution of the ODE~(\ref{ddotx0})
accurately tracks the evolution of the center of the DB soliton,
which progressively loses its contrast and eventually disappears in the condensate background, with the system converging to its ground state (see section 2).

\begin{figure}
\centering\includegraphics[scale=0.5]
{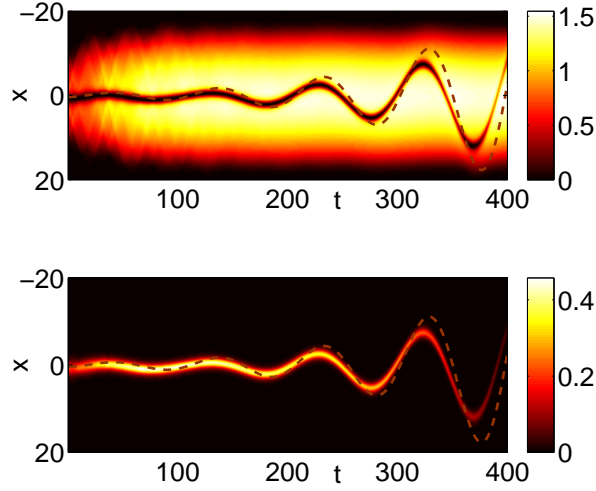}
\caption{(Color online)
Same as the bottom panel of Fig.~\ref{fig7}, but with an
initial density of the TF cloud at the trap center equal to $U_d(0) = 0.8$.}
\label{78}
\end{figure}

It is worth noting that in the results of Figs.~\ref{fig7} and
\ref{fig8} we have used, as initial condition, a TF cloud with a
density at the trap center equal to the chemical potential $\mu$
appearing in the DGPEs~(\ref{deq1})-(\ref{deq2}). Nevertheless, we
have also briefly studied a case where the density at $x=0$ of the TF
cloud was different from $\mu$. The evolution in such a far from
equilibrium scenario, is shown in Fig.~\ref{78}, where the parameters are
as in the subcritical
case of Fig.~\ref{fig7}, but with $U_d(0) = 0.8$. It is readily
observed that apart from the transient
 period towards equilibrium (i.e., when the density is rearranged so
 that it properly corresponds to the relevant value of the chemical potential $\mu=1.5$), the agreement between
analytical and numerical results
 is fairly good. I.e., the fast scale of the background relaxation
 does not substantially affect the evolution of the DB wave on the
slower time scale of the oscillatory decay of the latter.

\section{Two Dark-Bright Soliton States}

We now focus on the study of DB soliton ``molecules'' composed by two-DB-soliton states.
Let us first consider the homogeneous case ($\Omega=0$), and use the following ansatz to describe
a two-DB-soliton state
composed by a pair of two equal-amplitude, oppositely located (at $x=\pm x_0$) single DB solitons:
%
\begin{eqnarray}
\psi_1(x,t)
&=&      \left(\cos\phi\tanh X_{-}+i\sin\phi \right)
\left(\cos\phi\tanh X_{+}-i\sin\phi\right),
\label{eq15}
\\[1.0ex]
\psi_2(x,t)
&=& \eta\, {\rm sech} X_{-}\, {\rm e}^{i\left[kx+\theta(t)
\right]}
+\eta\, {\rm sech} X_{+}\, {\rm e}^{i\left[-kx+\theta(t)
\right]}\,
{\rm e}^{i\Delta\theta},
\label{eq16}
\end{eqnarray}
where $X_{\pm} = D\left(x \pm x_0(t)\right)$, $2x_0$ is the relative distance between the two solitons, and $\Delta\theta$ is the relative phase between the two bright solitons (assumed to be constant); below we will consider both the out-of-phase case, with $\Delta\theta=\pi$, as well as the in-phase case, corresponding to $\Delta\theta=0$.
Note that, similarly to the case of a single-DB soliton, the number of atoms $N_b$ of the bright-soliton component in the above two-DB-soliton state can be used to connect the DB-soliton parameters; in particular, if the two DB solitons are well-separated then
%
%
$N_b$ is approximately twice as large compared to the result of Eq.~(\ref{nb}), namely, $N_b \approx 4\eta^2 \sqrt{\mu}/D$.
%

As was recently shown in Ref.~\cite{ourrecent},
at zero temperature (i.e., $\gamma_d=\gamma_b=0$), the
evolution equation for the DB soliton center (for $\Omega=0$) reads:
%
\begin{eqnarray}
\ddot{x}_0 &=& F_{\rm int},
\label{eqmot} \\[1.0ex]
F_{\rm int}& \equiv &F_{\rm DD}+F_{\rm BB}+2F_{\rm DB}.
\label{Fint}
\end{eqnarray}
In the above equations, $F_{\rm int}$ is the interaction force between the two DB solitons,
which consists of
three different components: the interaction forces
$F_{\rm DD}$ and $F_{\rm BB}$ between the two dark and two bright components,
respectively, as well as the
interaction force $F_{\rm DB}$ of the dark soliton of the
one soliton pair with the bright soliton of the other pair (and vice-versa).
These forces depend on the soliton coordinate $x_0$, as well as on the DB soliton parameters, as follows
\cite{ourrecent}:
\begin{eqnarray}
F_{\rm DD}&=&\frac{1}{\chi_0}\left[\frac{1}{3}(544-352D_{eq}^2) +128D_{eq}\left( D_{eq}^2-1 \right)x_0 \right]
{\rm e}^{-4D_{eq}x_0},
\label{fdd} \\[2.0ex]
F_{\rm BB}&=&\frac{\chi}{\chi_0} \Big( -6D_{eq}+4D_{eq}^2x_0-2\chi \Big)
D_{eq}^2\cos\Delta\theta{\rm e}^{-2D_{eq}x_0} \nonumber\\
&+& \frac{\chi^2}{\chi_0} \Big[ \left(1+2\cos^2\Delta\theta\right)\left(-8D_{eq}x_0+6\right)\Big]
D_{eq}^2{\rm e}^{-4D_{eq}x_0},
\label{fbb} \\[2.0ex]
F_{\rm DB}&=&\frac{\chi}{\chi_0}\Big( 8D_{eq}\cos\Delta\theta \Big){\rm e}^{-2D_{eq}x_0}
-\frac{\chi}{\chi_0}\Big(\frac{208}{3}-64D_{eq}x_0 \Big) D_{eq}{\rm e}^{-4D_{eq}x_0},
\label{fdb}
\end{eqnarray}
where we have assumed that $\dot{D}(t)\approx 0$ and, thus, $D(t)\approx D_{eq}$.

Next, let us consider the case of two DB-solitons in the presence of the
harmonic trap. Then, each of the two solitons
is subject to two forces: (a) the restoring force of the trap, $F_{\rm tr}$ [in the case of
a single DB-soliton, this force induces an in-trap oscillation with a
frequency $\omega_{\rm osc}$ ---see Eq.~(\ref{omega})], and (b) the pairwise
interaction force $F_{\rm int}$ [cf.~Eq.~(\ref{Fint})] with other dark-bright
solitons. Thus, taking into regard that
$F_{\rm tr}=-\omega_{\rm osc}^2 x_0$, one may write the effective equation of
motion for the center $x_0$ of a two-DB-soliton state as follows:
\begin{equation}
\ddot{x}_0 = F_{\rm tr} + F_{\rm int}.
\label{eqmot2}
\end{equation}

\begin{figure}
\centering\includegraphics[scale=0.45]
{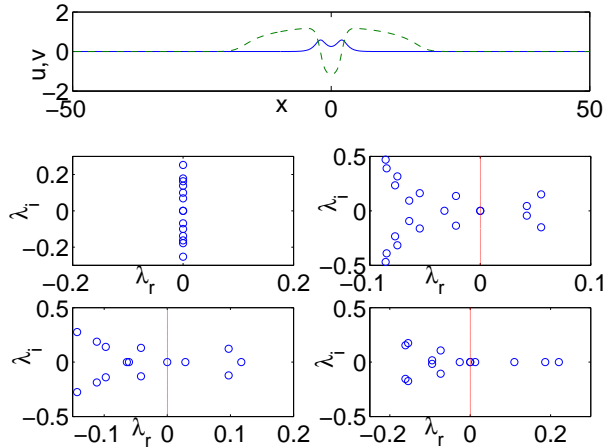}
\caption{(Color online)
The top panel depicts the stationary solution for an in-phase two-DB-soliton state
for $\mu=1.5$,
$|\Delta|=0.6$ and $\Omega=0.1$. The dark (bright) components
are shown by the dashed green (solid blue) lines.
The middle and bottom panels are four spectral planes, corresponding to different
values of $\gamma$, for the two DB-solitons in an in-phase, stationary configuration: in the
middle left panel $\gamma=0$ (the zero-temperature Hamiltonian case), in the middle
right $\gamma=0.1$, in the bottom left $\gamma=0.2$, and in the bottom right
$\gamma=0.4$.}
\label{fig3}
\end{figure}

\begin{figure}
\centering\includegraphics[scale=0.45]
{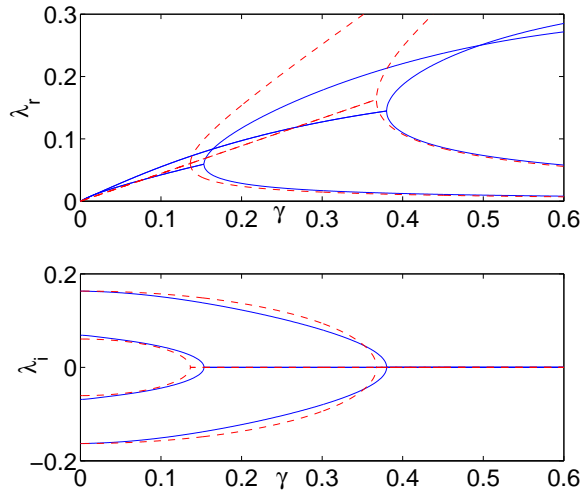}
\caption{(Color online)
The real (top panel) and imaginary (bottom panel) part of the two
anomalous mode eigenvalues for a two in-phase DB soliton state.
Both modes have  complex, for $\gamma\neq0$, and both pairs eventually collide and give rise to
exponential instabilities through eigenvalues on the real axis. Solid (blue) lines yield the
 numerical results, while the dashed (red) lines provide
the corresponding theoretical predictions of Eqs.~(\ref{frequency})-(\ref{two_db_frequency}).}
\label{fig4}
\end{figure}

In order to
complete the consideration of the case at hand, we will finally
study the finite-temperature effect on a two
DB-soliton state in the trap.
To do so, we will combine the thermal effect on each DB soliton in the trap, represented by Eq.~(\ref{ddotx0}),
and interaction effects included in Eq.~(\ref{eqmot}). This way, we may use the following approximation to describe
the motion of the centers of the two DB solitons:
%
\begin{eqnarray}
\ddot{x}_0-a\dot{x}_0-\left(F_{\rm tr} + F_{\rm int}\right) &=& 0.
\label{frequency}
\end{eqnarray}
The equilibrium points $x_{\rm eq}$, can easily be found as solutions of the
transcendental equation resulting from Eq.~(\ref{frequency}) letting $\dot{x}_0=\ddot{x}_0=0$ in both the in- and out-of-phase cases. To study the stability of these equilibrium points in the framework of
Eq.~(\ref{frequency}), we
use the ansatz $x_0(t)=x_{\rm eq}+\delta(t)$, and obtain a linear equation for the
small-amplitude perturbation $\delta(t)$, namely:
$\ddot{\delta}-a\dot{\delta}+\omega_1^2\delta=0$, where the frequency $\omega_1$ is given by,
\begin{eqnarray}
\omega_1^2 &=& \omega_{\rm osc}^2+\omega_{0}^2, \qquad
\omega_0^2= - \frac{\partial F_{\rm int}}{\partial x_0} \bigg|_{x_0 = x_{\rm eq}},
\label{two_db_frequency}
\end{eqnarray}
where
$\omega_{\rm osc}^2$ and $a$ are respectively
given by Eq.~(\ref{omega}) and
Eq.~(\ref{aa}).

\begin{figure}
\centering\includegraphics[scale=0.47]
{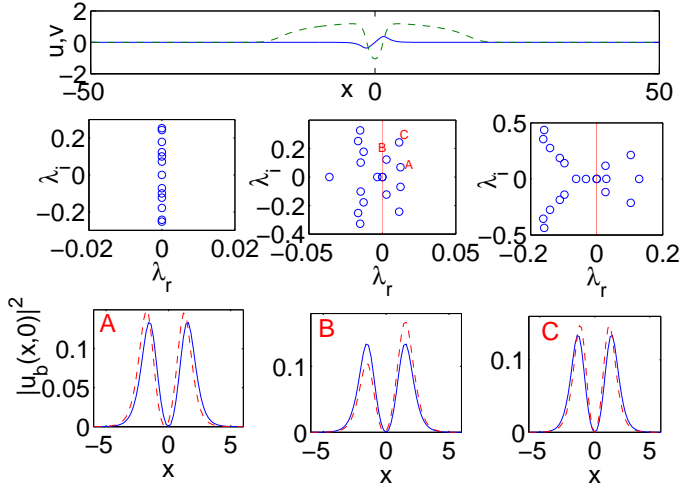}
\caption{(Color online)
The top panel depicts the stationary solution for two DB-solitons in an out-of-phase configuration, with parameters $\mu=1.5$, $|\Delta|=0.6$
and $\Omega=0.1$. The dark (bright) soliton components
are shown by the dashed green (solid blue) lines.
The middle panels show three spectral planes, corresponding to different
values of $\gamma$,
namely from left to right we have $\gamma=0$ (the zero-temperature
Hamiltonian case), $\gamma=0.02$ and $\gamma=0.2$ respectively.
In the bottom panel, we compare the bright soliton component's stationary solution (solid blue line), against the perturbed states (dashed red line) obtained
by adding to it the respective BdG eigenfunctions. More specifically
``A'', ``B'' and ``C'' correspond to the eigenfunctions of the three anomalous modes' eigenvalues in ascending order (see text).}
\label{fig5}
\end{figure}

\begin{figure}
\centering\includegraphics[scale=0.47]
{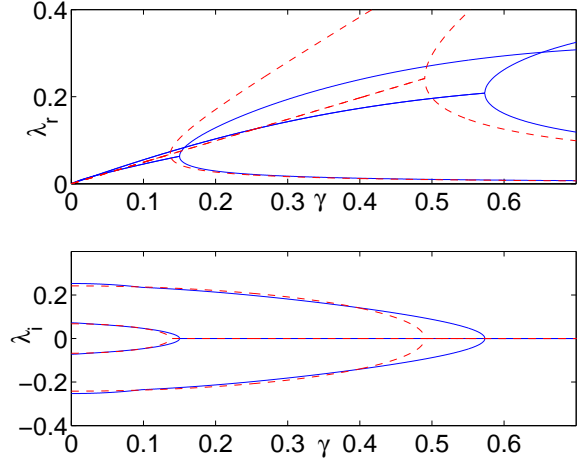}
\caption{(Color online)
The real part (top panel) and imaginary (bottom panel) part of the two
anomalous mode eigenvalues
(corresponding to ``A" and ``C" in the middle panel of Fig~\ref{fig5})
for an out of phase two DB soliton state. Both modes lead to Hopf bifurcations,
for $\gamma\neq0$, and both pairs eventually collide and give rise to exponential
instabilities through eigenvalues on the real axis. The solid (blue) lines
depict numerical results, while the dashed lines provide the corresponding theoretical predictions of Eqs.~(\ref{frequency})-(\ref{two_db_frequency}).}
\label{fig6}
\end{figure}

We now test the relevant predictions against BdG simulations, first for the in-phase case
in Figs.~\ref{fig3}-\ref{fig4} and then for the out-of-phase
case in Figs.~\ref{fig5}-\ref{fig6}.
As
expected, in the case of the in-phase configuration the BdG analysis reveals the existence of two anomalous modes: the one with the smaller (larger) eigenvalue---in the zero-temperature case---corresponds to an in- (out-of-) phase motion of the two DB solitons, similarly to the case of a two-dark-soliton state in a single-component BEC \cite{andreas,dcdss}.
These
two anomalous mode pairs
lead to complex eigenfrequencies for $\gamma \neq 0$, and the two-DB-soliton state performs oscillations of growing amplitude.
Similarly to the case of the single DB soliton, as $\gamma$ is increased
these pairs collide pairwise on the real axis in two critical
points, namely $\gamma_1 \approx 0.153$ and $\gamma_2 \approx 0.38$.
Beyond the second critical point $\gamma_2$, the
growth of the trajectory of the DB soliton center
becomes purely exponential. The theoretical
approximation of the relevant complex (and subsequently real)
eigenvalues depicted
by dashed line in Fig.~\ref{fig4}
is again fairly accurate, becoming progressively worse as $\gamma$ increases.

A similar phenomenology arises in the case of out-of-phase two-DB-soliton states, as shown in  Figs.~\ref{fig5}-\ref{fig6}. However, there exists a rather nontrivial twist in comparison to the previous
case. In particular, a third pair of complex eigenvalues emerges due
to the fact that a third anomalous mode exists for $\gamma=0$. This mode is no longer a {\it translational} one associated with the in- or out-of-phase motion of the two soliton centers (as before and as shown
in the bottom left and bottom right eigenmodes of Fig.~\ref{fig5}). It is instead a
mode associated with the $\pi$ relative phase of the peaks:
if we add the eigenvector of this unstable (for $\gamma \neq 0$)
mode to the two-DB-soliton out-of-phase solution, we observe that while
the center location of the state remains intact, the relative
heights of the two
solitons are affected, leading to a symmetry
breaking of the configuration. We will not consider this
unstable mode further since its induced instability is
weaker than those of the (in-phase and out-of-phase) translations.
Nevertheless, we note that all three pairs of modes eventually collide
on the real axis, eventually leading to pairs of purely real eigenvalues.

\begin{figure}
\centering\includegraphics[scale=0.5]
{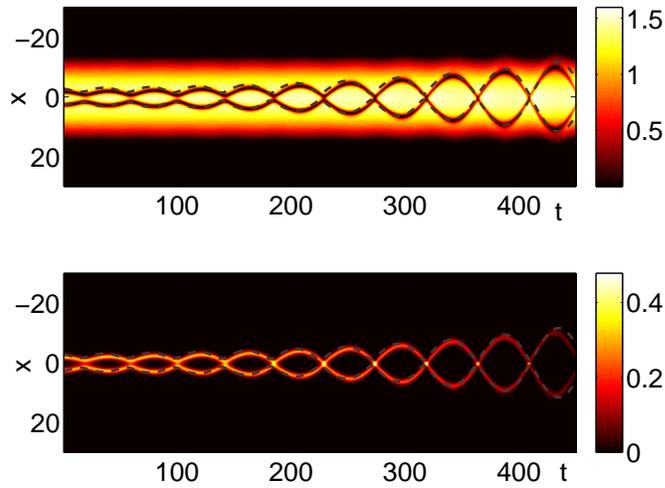}
\caption{(Color online)
Contour plot showing the space-time evolution of the density in the two
dark-bright soliton in-phase state.
The top panel represents the dark solitons and the bottom one the bright
solitons, with $\gamma=0.01$. The solitons
are initially placed at $x_1=2.75$ and $x_2=-2.75$. The dashed line represents the result obtained by
numerical solution of Eq.~(\ref{frequency}).}
\label{fig9}
\end{figure}

\begin{figure}
\centering\includegraphics[scale=0.5]{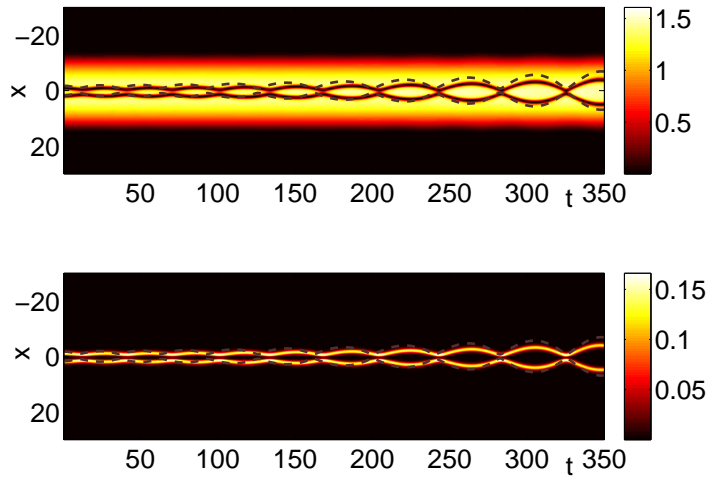}
\caption{(Color online)
Contour plot showing the space-time evolution of the density in the two dark-bright soliton out-of-phase state.
The representation of the solitons and the value of $\gamma$ is the
same as in Fig.~\ref{fig9}. The solitons
are initially placed at $x_1=1.76$ and $x_2=-1.76$. The dashed line
represents the result obtained by
numerical solution of Eq.~(\ref{frequency}).}
\label{fig10}
\end{figure}

Finally, we turn to direct numerical simulations for both the in-phase
two-DB-soliton state in Fig.~\ref{fig9}, and for the out-of-phase two-DB
state in Fig.~\ref{fig10}. In both cases,
we show only the low-$\gamma$, oscillatory growth (subcritical)
regime. Despite the complexity of the resulting system and of the DB soliton interactions, it can
still be clearly observed that the ODE~(\ref{frequency})
can be used to capture fairly accurately the relevant dynamics even for the long time evolutions considered
in these figures. Here, it should be mentioned that the temperature-induced dissipation results in an interesting effect. Particularly, as observed in Figs.~\ref{fig9} and \ref{fig10}, for short times, the individual DB solitons clearly behave like repelling particles, which can always be characterized by two individual density minima --- even at the collision point. Nevertheless, for longer times, the nature of their interaction changes: due to dissipation, they gain kinetic energy and completely overlap at the collision point.


\section{Conclusions}

In the present work, we presented a systematic analysis of a prototypical model (the so-called dissipative Gross-Pitaevskii equation) incorporating the effects of temperature on the dynamics of dark-bright (DB) solitons. This was done both in the
case of a single DB soliton, as well as in the case of DB soliton ``molecules'', composed by multiple (in- or out-of-phase) DB solitons.

We have developed a perturbation theory for the two-component system to analytically show the following: similarly to dark solitons, dark-bright ones execute anti-damped oscillations of growing amplitude for sufficiently low temperatures, while if the relevant parameter becomes sufficiently large, then the decay of the contrast of the solitons (and their disappearance in the background) becomes exponential.

A fundamental effect revealed by our analysis is that the presence of the bright (``filling'') component hinders the temperature-induced dissipation associated with the dark soliton, and offers a significant partial stabilization (i.e., a significantly longer lifetime) to the corresponding symbiotic DB soliton structure, in comparison to its ``bare'' dark soliton counterpart. The above effect relies on the fact that the critical value of the relevant parameter (labeling the different damping regimes) is increased, while the instability growth rate is decreased, for the DB solitons.
Similar conclusions were reached in the case of two dark-bright entities, with the added twist that their relative phase may introduce (in the out-of-phase case) additional anomalous modes and instability sources in the system. The latter are not associated with in- or out-of-phase translational motion of the solitons but rather with a symmetry-breaking in their relative amplitudes.

As concerns the relevance of our findings with pertinent experimental efforts we note the following. First, all relevant recent experiments for dark and dark-bright solitons were conducted at extremely low temperatures, aiming to minimize corresponding anti-damping effects. In our setting this corresponds to subcritical dynamics of small $\gamma$. Nevertheless, we believe that our findings may be relevant to future experiments exploring in more detail finite-temperature effects (see Supplemental Material of Ref.~\cite{pea}).

A natural direction to extend the present studies is to consider the
higher dimensional setting of vortices~\cite{jpb} and of their
two-component generalizations, namely the vortex-bright solitons~\cite{kody}.
Understanding the thermally induced dynamics and the modifications
of the corresponding precessional motion, especially in the presence
of multiple coherent structures would constitute an interesting topic
for future study. On the other hand, it would certainly be relevant
to extend the present studies to more complex models that
provide coupled dynamical equations for the condensate and the
thermal cloud~\cite{npprev} (rather than use a single equation directly
incorporating the effects of the thermal cloud on the condensate without the
possibility of ``feedback''). Such studies are in progress and pertinent results will be reported in future
publications.

\section*{Acknowledgments} P.G.K. gratefully acknowledges support
from the National Science Foundation through grants DMS-0806763
and CMMI-1000337, as well as from the Alexander von Humboldt Foundation
and the Alexander S. Onassis Public Benefit Foundation.
The work of D.J.F. was partially supported by the Special Account for Research Grants
of the University of Athens.

\vspace{1cm}



\begin{thebibliography}{99}

\bibitem{book2a} C. J. Pethick and H. Smith,
{\it Bose-Einstein condensation in dilute gases}, Cambridge University Press (Cambridge, 2002).


\bibitem{book2} L. P. Pitaevskii and S. Stringari, {\it Bose-Einstein Condensation}
(Oxford University Press, Oxford, 2003).


\bibitem{BECBOOK} P. G.~Kevrekidis, D. J.~Frantzeskakis, and R.~Carretero-Gonz\'alez (eds.)
{\it Emergent Nonlinear Phenomena in Bose-Einstein Condensates: Theory and Experiment}
(Springer Series on Atomic, Optical, and Plasma Physics, Vol.~\textbf{45}, 2008).

\bibitem{rab} F. Kh. Abdullaev, A. Gammal, A. M. Kamchatnov, and L. Tomio, Int. J. Mod. Phys. B {\bf 19}, 3415 (2005).

\bibitem{revnonlin} R. Carretero-Gonz\'alez, D. J. Frantzeskakis, and P. G. Kevrekidis,
Nonlinearity \textbf{21}, R139 (2008).

\bibitem{djf} D. J. Frantzeskakis, J. Phys. A: Math. Theor. {\bf 43}, 213001 (2010).

\bibitem{binary} C. J. Myatt, E. A. Burt, R. W. Ghrist, E. A. Cornell, and C. E. Wieman,
Phys. Rev. Lett. {\bf 78}, 586 (1997);
D. S. Hall, M. R. Matthews, J. R. Ensher, C. E. Wieman, and E. A. Cornell,
Phys. Rev. Lett. {\bf 81}, 1539 (1998).

\bibitem{mertes} K. M. Mertes, J. Merrill, R. Carretero-Gonz\'alez, D. J. Frantzeskakis,
P. G. Kevrekidis, and D. S. Hall, Phys. Rev. Lett. {\bf 99}, 190402 (2007).

\bibitem{hamburg} C. Becker, S. Stellmer, P. Soltan-Panahi, S. D\"{o}rscher, M. Baumert,
E.-M. Richter, J. Kronj\"{a}ger, K. Bongs, and K. Sengstock, Nature Phys. {\bf 4}, 496 (2008).

\bibitem{pea} C. Hamner, J. J. Chang, P. Engels, and M. A. Hoefer, Phys. Rev. Lett. {\bf 106}, 065302 (2011).

\bibitem{peb} S. Middelkamp, J. J. Chang, C. Hamner, R. Carretero-Gonz\'alez, P. G. Kevrekidis,
V. Achilleos, D. J. Frantzeskakis, P. Schmelcher, and P. Engels,
Phys. Lett. A {\bf 375}, 642 (2011);
M. A. Hoefer, C. Hamner, J.J. Chang, P. Engels, arXiv:1007.4947.

\bibitem{BA01} Th. Busch and J. R. Anglin, Phys. Rev. Lett. {\bf 87}, 010401 (2001).

\bibitem{the} D. Schumayer and B. Apagyi, Phys. Rev. A {\bf 69}, 043620 (2004);
K. Kasamatsu and M. Tsubota, Phys. Rev. A {\bf 74}, 013617 (2006);
X. X. Liu, H. Pu, B. Xiong, W. M. Liu, and J. B. Gong, Phys. Rev. A {\bf 79}, 013423 (2009);
H. Li, D. N. Wang, and Y. S. Cheng, Chaos, Solitons and Fractals {\bf 39}, 1988 (2009);
S. Rajendran, P. Muruganandam, and M. Lakshmanan, J. Phys. B: At. Mol. Opt. Phys. {\bf 42}, 145307 (2009);
V. A. Brazhnyi and V. M. P\'{e}rez-Garc\'{i}a, arXiv:1004.3672.

\bibitem{azucena} A. {\'A}lvarez, J. Cuevas, F.R. Romero and P.G. Kevrekidis, Physica D {\bf 240}, 767 (2011).

\bibitem{shl1} P. O. Fedichev, A. E. Muryshev, and G. V. Shlyapnikov, Phys. Rev. A {\bf 60}, 3220 (1999).

\bibitem{shl2} A. Muryshev, G. V. Shlyapnikov, W. Ertmer, K. Sengstock, and M. Lewenstein,
Phys. Rev. Lett. {\bf 89}, 110401 (2002).

\bibitem{us} S. P. Cockburn, H. E. Nistazakis, T. P. Horikis, P. G. Kevrekidis,
N. P. Proukakis, and D. J. Frantzeskakis, Phys. Rev. Lett. {\bf 104}, 174101 (2010).

\bibitem{ft} N. P. Proukakis, N. G. Parker, C. F. Barenghi, and C. S. Adams, Phys. Rev. Lett. {\bf 93}, 130408 (2004);
B. Jackson, N. P. Proukakis, and C. F. Barenghi, Phys, Rev. A {\bf 75}, 051601 (2007);
B. Jackson, C. F. Barenghi, and N. P. Proukakis, J. Low Temp. Phys. {\bf 148}, 387 (2007);
W. H. Zurek, Phys. Rev. Lett. {\bf 102}, 105702 (2009);
A. D. Martin and J. Ruostekoski, Phys. Rev. Lett. {\bf 104}, 194102 (2010);
A. D. Martin and J. Ruostekoski, New J. Phys. {\bf 12}, 055018 (2010);
B. Damski and W. H. Zurek, Phys. Rev. Lett. {\bf 104}, 160404 (2010).

\bibitem{gk} D. M. Gangardt and A. Kamenev, Phys. Rev. Lett. {\bf 104}, 190402 (2010).

\bibitem{ashton} K. J. Wright and A. S. Bradley, arXiv:1104.2691.

\bibitem{han1} S. Burger, K. Bongs, S. Dettmer, W. Ertmer, K. Sengstock, A. Sanpera,
G. V. Shlyapnikov, and M. Lewenstein, Phys. Rev. Lett. {\bf 83}, 5198 (1999).

\bibitem{nist} J. Denschlag, J. E. Simsarian, D. L. Feder, C. W. Clark, L. A. Collins, J. Cubizolles, L. Deng,
E. W. Hagley, K. Helmerson, W. P. Reinhardt, S. L. Rolston, B. I. Schneider, and W. D. Phillips,
Science {\bf 287}, 97 (2000).

\bibitem{han2} K. Bongs, S. Burger, S. Dettmer, D. Hellweg, J. Arlt, W. Ertmer, and K. Sengstock,
C. R.  Acad. Sci. Paris {\bf 2}, 671 (2001).

\bibitem{kip} A. Weller, J. P. Ronzheimer, C. Gross, J. Esteve, M. K. Oberthaler,
D. J. Frantzeskakis, G. Theocharis, and P. G. Kevrekidis,
Phys. Rev. Lett. {\bf 101} (2008), 130401.

\bibitem{hambcol} S. Stellmer, C. Becker, P. Soltan-Panahi, E.-M. Richter, S. D\"{o}rscher, M. Baumert,
J. Kronj\"{a}ger, K. Bongs, and K. Sengstock,
Phys. Rev. Lett. {\bf 101}, 120406 (2008).

\bibitem{andreas} G. Theocharis, A. Weller, J. P. Ronzheimer, C. Gross,
M. K. Oberthaler, P. G. Kevrekidis, and D. J. Frantzeskakis,
Phys. Rev. A {\bf 81}, 063604 (2010).


\bibitem{lp} L. P. Pitaevskii, Zh. Eksp. Teor. Fiz. {\bf 35}, 408 (1958)
[Sov. Phys. JETP {\bf 35}, 282 (1959)].

\bibitem{npprev} B. Jackson and N. P. Proukakis,
J. Phys. B: At. Mol. Opt. Phys. {\bf 41}, 203002 (2008).

\bibitem{stoch} S. P. Cockburn and N. P. Proukakis,
Laser Phys. {\bf 19}, 558 (2009).

\bibitem{dcdss} P. G. Kevrekidis and D.J. Frantzeskakis,
Discr. Cont. Dyn. Sys. S {\bf 4}, 1199 (2011).

\bibitem{ourrecent} D. Yan, J. J. Chang, C. Hamner, P. G. Kevrekidis, P. Engels, V. Achilleos,
D. J. Frantzeskakis, R. Carretero-Gonzalez, P. Schmelcher, arXiv:1104.4359.


\bibitem{st1} H. T. C. Stoof and M. J. Bijlsma, J. Low Temp. Phys. {\bf 124}, 431 (2001);
R. A. Duine and H. T. C. Stoof, Phys. Rev. A {\bf 65}, 013603 (2001).

\bibitem{st2} M. J. Davis, S. A. Morgan, and K. Burnett, Phys. Rev. Lett. {\bf 87}, 160402 (2001);
A. Sinatra, C. Lobo, and Y. Castin, Phys. Rev. Lett. {\bf 87}, 210404 (2001).






\bibitem{jap} H. Takeuchi, N. Suzuki, K. Kasamatsu, H. Saito, and M. Tsubota, Phys. Rev. B {\bf 81}, 094517 (2010).

\bibitem{cockburn} S. P. Cockburn, ``Bose Gases in and out of equilibrium within the
stochastic Gross–Pitaevskii equation'', Ph.D. Thesis (Newcastle University, 2010).

\bibitem{bur} S. Choi, S. A. Morgan, and K. Burnett, Phys. Rev. A {\bf 57}, 4057 (1998).

\bibitem{sand} T. Kapitula, P. G. Kevrekidis and B. Sandstede, Physica D {\bf 195}, 263 (2004).

\bibitem{jpb} S. Middelkamp, P.G. Kevrekidis, D.J. Frantzeskakis,
R. Carretero-Gonz{\'a}lez and P. Schmelcher,
J. Phys. B {\bf 43}, 155303 (2010).

\bibitem{kody}  K.J.H. Law, P.G. Kevrekidis and L.S. Tuckerman,
Phys. Rev. Lett. {\bf 105}, 160405 (2010).

\end{thebibliography}
\end{document}